# Immunopathogenesis in Psoriasis through a Density-type Mathematical Model

BIPLAB CHATTOPADHYAY* and NIRMALENDU HUI+
Department of Physics
Acharya B. N. Seal College
Cooch Behar 736101, West Bengal
INDIA
*bchattopa@bsnl.in ; +nirmal_hui@yahoo.com

*Abstract:* - Disease psoriasis occurs as chronic inflammation of skin and appears as scaly red lesions on skin surface. Advent of several immunosuppressive drugs established that the disease stems from immuno-pathogenic disorder in human blood. Cell biological as well as clinical research on the disease reveals that the helper T-cells and other Leucocytes, responsible for human immunity, may lead to psoriasis pathogenesis if produced in plenty at locations close to the dermal region. Research findings also showed that a complex, self-sustaining (cytokine and related) proteins network play important role in disease maturation by actually leading to a huge proliferation of epidermal keratinocytes. Disease pathogenesis is identified with such hyperproliferation leading to flaking of skin surface (psoriatic plaques). An excessive generation of nitric oxide by proliferated keratinocytes, through a complex chain of bio-chemical events, is causal to the scaliness of psoriatic plaques. Considering these immunopathogenic mechanisms, we propose and analyse a mathematical time differential model for the disease psoriasis. Outcomes of analysis are consistent with existing cell biological and clinical findings with some new predictions which could be tested further.

## 1 Introduction

The terminology psoriasis is inherently linked to the connotation 'soreness of skin' which essentially implies a disease occurring as chronic inflammation of skin. Actually, the disease psoriasis is manifested as scaly red lesions on the skin surface termed as plaques. Presently, more than 3 % of the world population is affected by the disease and there is significant morbidity. As of now, exact mechanism leading to the disease is not precisely understood. Consequently, any specific and systematic therapy to completely cure the disease has not yet been evolved. Under the present circumstances, it is generally believed that the exact causes for infliction of the disease psoriasis are yet to be fully understood and hence psoriasis is incurable. However, clinical and cell biological research, till date, point towards various genetic, environmental and immunological factors contributing to pathogenesis of the disease [1-4].

It is being evidenced by clinical research that psoriasis is caused due to a breakdown in the human (mammalian) immune system where the disease fighting immune cells (T-cells) virtually attack the healthy skin cells due to some erroneous signaling [5]. This conclusive statement is emboldened by the fact that, immunosuppressive drugs like cyclosporine A, which inhibits T-cell proliferation and cytokine production, have been observed effective in treating psoriatic lesions [6]. Further, trials with anti-CD4 monoclonal antibodies and IL-2 toxin conjugate on psoriatic patients have found remedial response to psoriatic plaques [7, 8]. Thus, psoriasis is being characterized as an autoimmune disease and is probably the most prevalent in the category [9-11].

Disease maturation in psoriasis is observed to be led by hyperproliferation of epidermal keratinocytes which produces thickening or flaking of epidermis [12]. Within the psoriatic skin lesions, a sizeable influx of myeloid dendritic cells takes place [13] leading to substantial generation of cytokine $TNF-\alpha$ and thus causing inflammation of epidermis [14]. In addition, high levels of cytokines (proteins) such as $IFN-\gamma$, IL-17, $TGF-\beta$ have been found in psoriatic skin as in other inflammatory diseases [15-18]. Further, epidermal keratinocytes in psoriatic plaques produce the enzyme *i*-NOS (inducible nitric oxide synthase) which gives way to massive nitric oxide production [19, 20]. In fact, nitric oxide production in psoriatic epidermis seems to be ten times higher than that in non-psoriatic epidermis with normal level of keratinocytes [21]. This massive nitric oxide, produced through *i*-NOS pathway at the cost of hyperproliferated keratinocytes, induce over-sensitivity in the psoriatic plaques and hence their redness [22].

A complex cytokine network is closely linked to the process of pathogenesis in case of psoriasis. Actually, various cytokines, chemokines, growth factors, which are broadly clubbed in the protein category, are vital for the disease maturation [14-18, 23, 24]. Cytokines IFN$-\gamma$, TNF-$\alpha$ IL-2, which are secreted by T-helper 1 (Th-1) cells, are predominant in psoriasis, out of which IFN-$\gamma$ is produced by CD4+T and CD8+T cells. It has been observed that lesions can be induced by intradermal injection of IFN$-\gamma$ [15]. Cell biological research testifies that Leucocytes (immune cells) and keratinocytes can produce various cytokines and chemokines [12]. In fact, Th-1 cytokines IFN$-\gamma$ and TNF$-\alpha$, induce keratinocytes to further produce other cytokines like IL-6, IL-8, IL-12, and IL-18 [25-28]. A type of leukocyte, termed as dendritic cells, secretes the cytokine IL-23 which, in turn, stimulates T-helper 17 leucocytes to add to the production of other cytokines such as IL-17A, IL-17F, IL-6, IL-22 and TNF-$\alpha$ [17]. Various proteins, such as cytokines, chemokines, growth factors, get synergized mutually within themselves to further enhance their weight (density) in the blood and thereby increase influx of leucocytes in the epidermal skin as well as cause keratinocyte hyperproliferation [12]. Proteins also increase apoptotic resistance of keratinocytes. Increased influx of leucocytes in the skin gets induced by keratinocytes and contributes to further proliferation of epidermal keratinocytes [29]. The protein network (comprising of cytokines, chemokines, growth factors etc) in psoriasis is, thus, a self sustaining phenomenon which immensely influence and regulate pathogenesis in psoriasis. As perceived by clinical and cell-biological research, interplay between these proteins in a fairly complex network can explain most of clinical features in psoriasis.

Information gathered from clinical and cell-biological investigations, as portrayed above, when sequenced systematically, point towards a logical and formative depiction, of which a mathematical archetype can be framed. Framing of such archetype is termed as the process of mathematical modeling in general and for the present case it is for the disease psoriasis. In this communication, we wish to formulate, in detail, a density type mathematical model to represent immunopathogenesis for the disease psoriasis. We would also like to perform an in-depth analysis of the mathematical model, so formulated, to bring forth relevant and salient results and some predictions suggestive of cure from the disease, which could be tested further clinically.

The paper is organized as follows. In section-2 we describe, in detail, the process of framing the mathematical archetype for the disease psoriasis. With a suitable background-narration and having detailed the assumptions leading to the density type mathematical archetype of the disease psoriasis, theoretical analysis relating the stabilities of the model is being taken up in Section-3. In section-4, we take up the task for analyzing detailed of the model through the numerical avenue. Results from the numerical analysis have also been discussed in Section-4. Some concluding remarks including new predictions as well as future directions of work in the area, is being presented in Section-5.

## 2 Formulation of the mathematical archetype

Accumulating research results from the clinical and cell biological investigations, emergence of a mathematical model to represent immuno-pathogenesis in psoriasis can be asserted, albeit with logical and coherent sequencing of such results. The mathematical model in question is a set of time differentials of some associated variables representing population/mass densities of biological cell/molecules, with these differentials equated to significant cell-biological processes of associative relevance, involving a set of model parameters.

Based on clinical and cell-biological findings, a preliminary model was earlier proposed involving densities of T-cells, dendritic cells and keratinocytes as model variables [30-32]. This preliminary model, on being analyzed, signified disease pathogenesis in some region of parameter space, but the results were menacingly simple and abnormally linear in nature. A fundamental reason for such shortcomings has been the overlooking of important role of cytokines, chemokines, growth factors and such other proteins in the process of psoriasis pathogenesis. This realization of the significance of proteins in psoriasis, as depicted in the introduction part in this article, makes us to assert that proteins need direct inclusion in the mathematical archetype of psoriasis.

Taking leads from the clinical and cell biological rulings, as in the introduction, here we give a linguistic depiction of the mathematical model for psoriasis, narrating the processes that either enrich or degrade a model variable. Let us consider that, *leucocytes* ($L$), *proteins* ($P$) and $keratinocytes$ ($K$) represent three model variables. Cell-biological processes linked to time differential of Leucocytes are:

(a) An upstream influx of T-cells dendritic cells and other leucocytes in the affected region of epidermis.
(b) Various proteins contribute to leucocyte population,
(c) Loss of leucocytes through their contribution to proteins and through other natural processes.

Processes related to time variation of proteins are:

(a) Secretion of various proteins by leucocytes.
(b) Keratinocytes in conjugation with proteins contribute further to protein density.
(c) Loss of proteins through their giving in to leucocytes and to hyperproliferation of keratinocytes.

Time variation of keratinocytes is being associated to the following processes:

(a) Leucocyte stimulate keratinocytes to proliferate,
(b) Proteins directly yield to keratinocyte hyperproliferation.
(c) Loss of keratinocytes through its contribution to proteins.
(d) Loss of proliferated keratinocytes owing to nitric oxide production through i-NOS pathway.

Having stated the dominant cell-biological and allied phenomena linked to the three model variables $L, P$ and $K$ (in units of $mm^{-3}$), we proceed for narrating the customary assumptions to arrive at the basic model in mathematical connotation. Following are the assumption:

A1. We assume an influx of leucocytes (immune cells) at a constant rate '$a$' in the proximal spatial region of the dermal skin prospective to be affected by psoriasis. Predominantly the leucocytes, of T-cells and dendritic cells category, are prone for upstream influx.

A2. To cohere to the essence of A1, it is assumed that the relevant leucocytes are not reproduced on their own, by any mechanism or in any form.

A3. We assume that leucocytes, being stimulated, contribute to the production of proteins at a constant rate $\boldsymbol{\beta}$ ($\in R_+$). Various proteins such as cytokines, chemokines, growth factors etc., in turn, directly yield to the proliferation of leucocytes at the relevant region of space at a constant rate $\boldsymbol{\alpha}$ ($\in R_+$). Such proliferation of leucocytes is entirely due to the protein mass by stimulation.

A4. It is assumed that leucocytes, being induced by keratinocyte mass, further add to the growth of epidermal keratinocytes at the rate $\eta$ ($\in R_+$).

A5. We assume that various proteins directly yield to the hyperproliferation of keratinocyte mass at a rate $\eta'(\in R_+)$.

A6. Proliferated keratinocytes are assumed to have propotional interaction with the protein mass and such interactions lead to the enhancement of protein mass at a rate $\delta$ ($\in R_+$).

A7. A per capita removal of leucocytes at the rate $\mu$ ($\in R_+$) is assumed, which is attributed to the loss of leucocytes owing to natural processes and for their (leucocytes) contributing to the protein concentration as well as keratinocyte proliferation.

A8. A per capita removal of proteins at the rate $\lambda$ ($\in R_+$) is assumed on account of their contributing to leucocyte and keratinocyte concentrations.

A9. We assume removal of keratinocytes at a rate $\gamma$ ($\in R_+$), owing to their interaction with proteins and their eventual contribution to the protein mass.

A10. We assume a per capita loss of keratinocyte mass, at a constant rate $\boldsymbol{\lambda}'$ ($\in R_+$), due to its generating nitric oxide through $i$-NOS pathway [33, 34].

The above assumptions A1-A10, while considered together, lead to three time differential equations for the variables $L, P$ and $K$ as

$$\frac{dL}{dt} = a + \alpha P - \mu L$$

$$\frac{dP}{dt} = \beta L + \delta PK - \lambda P \qquad (1)$$

$$\frac{dK}{dt} = \eta LK + \eta' P - \gamma PK - \lambda' K$$

This set of three time differential equations (1), actually contribute to the mathematical archetype or model of the disease psoriasis and these model equations can be analyzed to understand the immunopathogenesis in psoriasis.

## 3 Analyses on Model Stability

By observation, it can be asserted that right hand side of the model equation (1) are smooth functions of variables $L, P, K$ and the various model parameters defined in the earlier section. While forming the model equation we have inherently assumed that the model variables $L, P, K$ and the model parameters are all non-negative and these assume values on the real lines. Thus it can be emphasized that the solutions of the model equations do have local existence in specified domains as well as they hold uniqueness and continuity properties in the positive octant of co-ordinate space.

The model equations (1) may further be analyzed to judge the existence of different equilibria which would signify the stability of the allied biological system. In general, the following equilibrium may exist on the different co-ordinates planes. These equilibria are: (i) $E_1(L_1^*,0,0)$ with $L_1^* = a/\mu$ and the parameter $a = 0$ is a possibility. (ii) $E_2(L_2^*, P_2^*, 0)$ with $L_2^* = \frac{a\lambda}{\mu\lambda - \alpha\beta}$ & $P_2^* = \frac{a\beta}{\mu\lambda - \alpha\beta}$ and (iii) $E^*(L^*, P^*, K^*)$ where the asymptotic stable values of variables can be obtained by solving the following equations

$a + \alpha P^* - \mu L^* = 0$
$\beta L^* + \delta P^* K^* - \lambda P^* = 0$ (2)
$\eta L^* K^* + \eta' P^* - \gamma P^* K^* - \lambda' K^* = 0$

Equations (2) can be solved for asymptotic stable values of model variable which turns out to be

$$P^* = \frac{\mu L^* - a}{\alpha} \quad (3.a)$$

$$K^* = \frac{\eta'(\mu L^* - a)}{\alpha\lambda' + \gamma(\mu L^* - a) - \alpha\eta L^*} \quad (3.b)$$

And $L*$ could be obtained from

$L^{*2}(xy + \mu^2 v) + L^*(xz + \lambda ay - 2\mu av) +$
$(\lambda az + a^2 v) = 0$ (4)

With

$x = (\alpha\beta - \lambda\mu), \quad y = (\gamma\mu - \alpha\eta), \quad z = (\alpha\lambda' - \gamma a),$
$v = \delta\eta'$

Equation (4) yields solutions for $L^*$ as

$L^* =$
$\frac{-(xz + \lambda ay - 2\mu av) \pm \sqrt{\{(xz + \lambda ay - 2\mu av)^2 - 4(xy + \mu^2 v)(\lambda az + a^2 v)\}}}{2(xy + \mu^2 v)}$

$L^*$ would assume values on real line provided

$(xz + \lambda ay - 2\mu av)^2 - 4(xy + \mu^2 v)(\lambda az + a^2 v)$
$\geq 0.$

Further analyzing the above conditional relation we obtained the following condition on the numerical values of parameter

$$\alpha\beta > \lambda\mu, \quad \lambda\eta > \beta\gamma \quad (5.a)$$

and

$$\alpha\lambda' > \gamma a, \quad a\eta > \mu\lambda' \quad (5.b)$$

Note that the last equilibrium $E^*(L^*, P^*, K^*)$ signifies the general stability of the asymptotic solutions and the allied parametric relations are useful in standardizing the model parameters. This $E^*$ is also termed as the interior equilibrium of the mathematical model and its existence is dependent on the imposing conditions, as equations (5) so derived, relating various model parameter.

Let us judge the stability of the general equilibrium $E^*(L^*, P^*, K^*)$ of the model under consideration which necessitates finding out the Jacobian matrix for the model [35]. For this purpose one needs to linearize model equations (1) around the equilibrium point by introducing a set of new variables which measure the deviation about the equilibrium. Thus we introduce

$$\begin{array}{l} X(t) = L(t) - L^* \\ Y(t) = P(t) - P^* \\ Z(t) = K(t) - K^* \end{array} \quad (6)$$

We then linearize the model equations (1) about the general equilibrium $E^*$ following standard procedure and we get the matrix equation

$$\begin{pmatrix} \dot{X} \\ \dot{Y} \\ \dot{Z} \end{pmatrix} = \begin{pmatrix} -\mu & \alpha & 0 \\ \beta & \delta K^* - \lambda & \delta \\ \eta K^* & \eta' - \gamma K^* & \eta L^* - \gamma P^* - \lambda' \end{pmatrix} \begin{pmatrix} X \\ Y \\ Z \end{pmatrix}$$

$$= \mathbf{J} \begin{pmatrix} X \\ Y \\ Z \end{pmatrix} \quad (7)$$

Where **J** is called the Jacobian Matrix.

Since we are interested about the stability of the general equilibrium $E^*(L^*, P^*, K^*)$ we apply the Routh-Hurwitz criterion [35] for the Jacobian in the above matrix equation to obtain the conditions

$$Tr \, \mathbf{J} < 0 \text{ and } det \, \mathbf{J} > 0. \quad (8)$$

Our calculations yield the values of trace and determinant of the Jacobian as,

$$Tr\ J = -(\mu + \lambda + \lambda') + \delta K^* + \eta L^* - \gamma P^* \quad (9)$$

and,

$$det\ J = \eta\lambda(\mu - \beta)L^* + (\eta'\mu\delta - \mu\lambda\gamma + \gamma\alpha\beta)P^* + \lambda'\mu\delta K^* - \mu\delta\eta L^*K^* + \alpha\eta\delta P^*K^* + \lambda'(\mu\lambda + \alpha\beta) \quad (10)$$

Further, we find that in our model as defined in equations (1), Routh-Hurwitz criterion is satisfied for the interior equilibrium $E^*$ and hence we may assert that the system is locally asymptotically stable around the general equilibrium $E^*(L^*, P^*, K^*)$.

## 4 Analyses In The Numerical Avenue

Numerical solution of the model equation (1) becomes a necessity in order for graphical illustration and visualization of the dynamical progression of the model outcomes. Detailed understanding of the model dynamics can be achieved through numerics of the model equations. Further, numerical solutions make it possible to study the effect of model parameters on model dynamics and lead to estimation of various thresholds existing within the model.

**Table 1**: Default values of Model Parameters

| Parameter | Definition | Default Values |
|---|---|---|
| $a$ | Rate of influx of Leucocytes | 5 $mm^{-3}$ $Day^{-1}$ |
| $\alpha$ | Rate of growth of Leucocytes out of Proteins | 0.03 $Day^{-1}$ |
| $\mu$ | Per capita removal of Leucocytes | 0.18 $Day^{-1}$ |
| $\beta$ | Growth rate of Proteins out of Leucocytes | 0.0005 $Day^{-1}$ |
| $\delta$ | Growth rate of Proteins by P & K mixing | 0.0001 $mm^3$ $Day^{-1}$ |
| $\lambda$ | Per capita removal of Proteins | 0.08 $Day^{-1}$ |
| $\eta$ | Growth rate of Keratinocytes by L & K | 0.011 $mm^3$ $Day^{-1}$ |
| $\eta'$ | Growth-rate of Keratinocytes due to P | 0.0005 $Day^{-1}$ |
| $\gamma$ | Rate of reduction of Keratinocytes by its interaction with Proteins | 0.0001 $mm^3$ $Day^{-1}$ |
| $\lambda'$ | Rate of removal of Keratinocytes for i-NOS & for giving in to Proteins | 0.06 $Day^{-1}$ |

In order to solve the model equations numerically one needs to assign default or standard numerical values to the set of model parameters and the same is done in Table 1. In the present case, standardization of parameters are done, predominantly by going through the accumulated clinical data on psoriasis and based on the so far achieved experimental results in the domain of cell-biological research on psoriasis. It is to be noted here that the potential tool of numerical solution has also been used extensively as one in many acceptable processes of standardization of model parameters. Actually, taking maiden leads from clinical and cell-biological research findings, some judicious or suitable values of model parameters are chosen and then all the model parameters are varied in wide ranges to get solutions of model variables from equations (1). Thus, continually judging the numerical solutions of model equations, the standard value of each parameter has been reached by repeated numerical trials. Imposing condition involving model parameters, as yielded by theoretical analysis on model (solution) stability have also been borne in mind during the process of standardization of default parameter values.

Considering default model parameters values as demonstrated in Table 1, model equations (1) are solved numerically. Results from such numerical solutions are explained through relevant planar viewgraphs termed as figures. Primarily, the time series solutions of model variables are analyzed to understand their dependences on various model parameters and then, asymptotic solutions of model variables are studied for changing values of parameters to bring out characteristic dependences of asymptotic solutions on various model parameters.

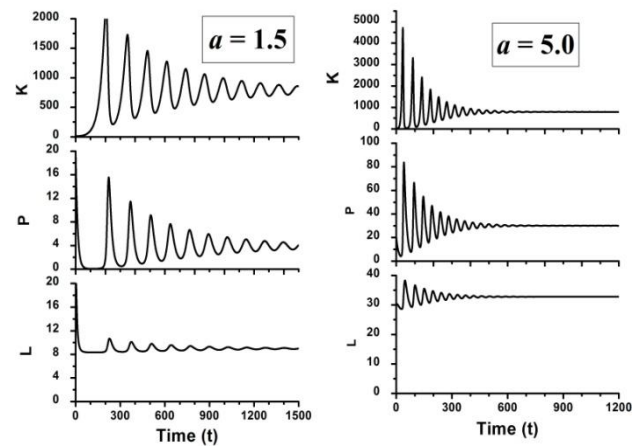

**Fig.1:** Time series solutions of model variables for two different values of *a,* as depicted.

In Fig.1, we plot time series solutions for variables *L*, *P* and *K* as functions of time (taken in days) for two different values of parameter $a$ denoting the rate of upstream influx of leucocytes. An increase in $a$ signifies considerable dying down of oscillations in $L(t)$, $P(t)$ and $K(t)$ as well as an enhancement of asymptotic values of solutions including that of asymptotic Keratinocyte density $K(t)$. Thus an increased $a$ seems to favour psoriasis pathogenesis and the reverse process, that is considerable lowering in $a$, implies a roll back from pathogenesis.

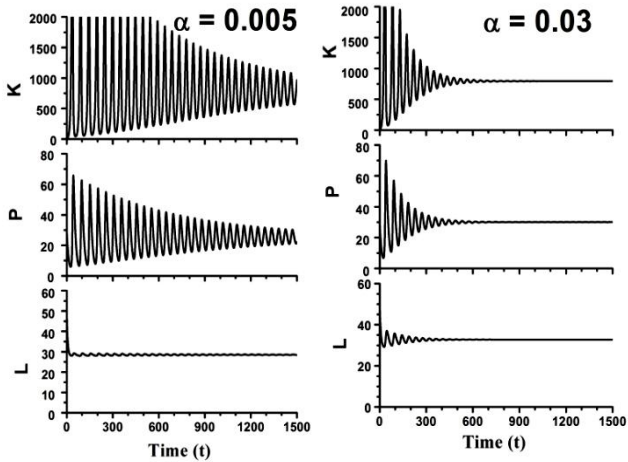

**Fig.2:** Time series solutions of model variables for two different values of $\alpha$, as given in the figure.

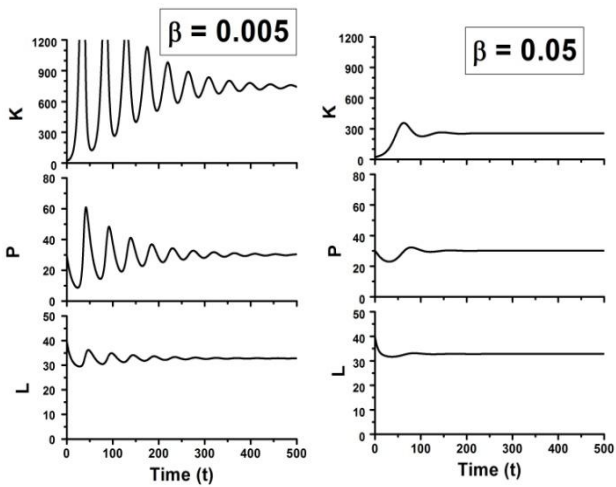

**Fig.3:** Time series solutions of model variables for two different values of model parameter $\beta$.

Solutions for leucocytes $(L)$, proteins $(P)$ and keratinocytes $(K)$ are plotted as a function of time (in days) for two different chosen values of the parameters $\alpha$, the growth rate of leucocytes owing to proteins, in Fig.2. We find here that for smaller $\alpha$, solutions remain oscillatory till a very large time, whereas, with higher $\alpha$ the solutions become single valued and stable much faster. It is apparent from the viewgraph that with $\alpha = 0.005$ $P(t)$ and $K(t)$ remains oscillatory and multi-valued even at a fairly large time $t = 1500$ days. But with an increase of $\alpha$ by an order of magnitude at 0.03, all three variables equilibrate to single valued solutions at around $t \sim 900$ days. Increase in $\alpha$ is accompanied by nominal increase in the asymptotic leucocytes' density.

Plot of $L(t)$, $P(t)$ and $K(t)$ solutions, for different $\beta$ denoting the rate of protein growth owing to lucocytes, has been included in Fig.3. Parameters other than $\beta$ are kept at their standardized values as in Table 1.

Increase in $\beta$ plays such that the single valuedness of solutions are reached considerably faster on time scale. Significant fall in the asymptotic stable value of keratinocyte density $(K)$, for increasing $\beta$, is note-worthy. Asymptotic values of leucocyte $(L)$ and protein $(P)$ masses remain almost insensitive to change in $\beta$, as observed in the plot.

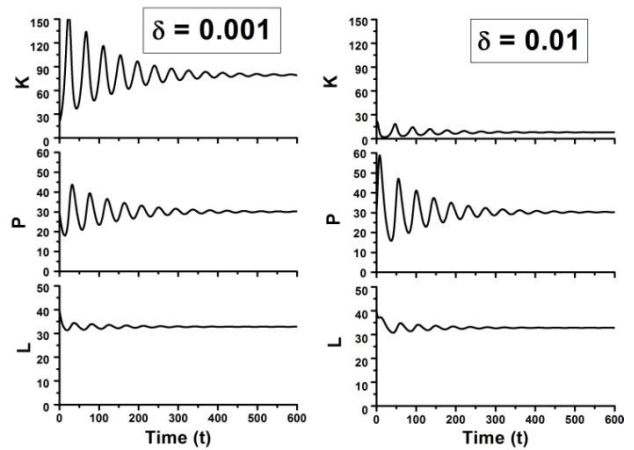

**Fig.4:** Solutions for model variables as function of time for two different values of model parameter $\delta$.

Fig.4 contains plots of time series solutions of $L$, $P$ and $K$ masses for two different values of the parameter $\delta$, representing growth rate of proteins owing to its mixing with keratinocyte mass. Here we find that all three solutions become single valued at a fairly smaller time $t \sim 500$ days when compared to the solutions in the earlier figures. We also observe that increase in $\delta$ makes the asymptotic keratinocyte density to degrade drastically and $P(t)$ oscillation amplitude is higher at small $t$.

In all the above figures, we presented time series solutions of model variables as functions of time taken in units of days. A keen observation reveals that asymptotically all the solutions become single valued and stable irrespective of the numerical values assigned to the parameters. It is also apparent that there are situations where asymptotic keratinocyte density stabilizes at high values signifying psoriasis pathogenesis. However, in such pathogenic solutions conditions, we always find that solutions equilibrate to single valued stable state at a large time $t \sim 1500 - 1700$ days. This phenomenon manifestly shows that disease maturation (or the process of pathogenesis) in psoriasis requires a long time which is compatible to those observed by the researches in the clinical domain.

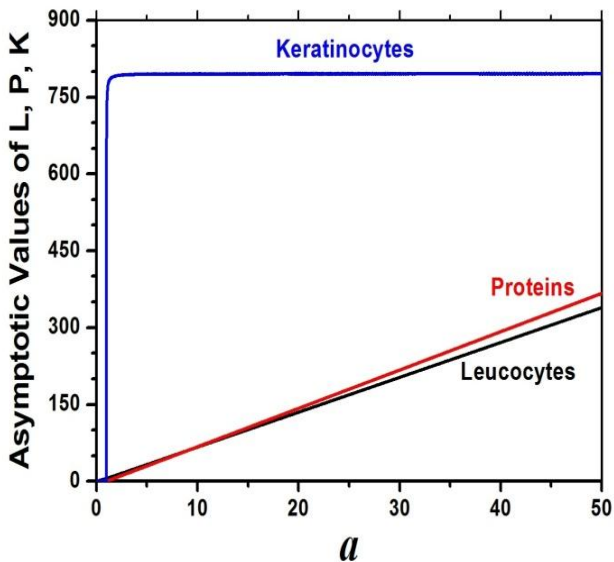

**Fig.5:** Phase-diagram depicting dependences of asymptotic stable solutions of model variables $(L^*, P^*, K^*)$ with varying model parameter $a$.

Next we would like to present numerical results pertaining to the phase diagram of model variable (solutions) as function of different model parameters. Here we actually produce plots of asymptotic (stable) solutions for model variables with varying model parameters. It should be noted that for solving differential model equations we have set the time-step at $\Delta t = 0.001$ and explored up to $t = 2000$ days so that the solutions are really asymptotic for all practical purposes.

Fig.5 represents phase diagram involving asymptotic stable densities of model variables $L^*$, $P^*$ and $K^*$ plotted as a function of the parameter $a$ while all other parameters are kept at their default values as in Table 1. We observe that fixed values of variables $L^* = L(t \to \infty)$, $P^* = P(t \to \infty)$ and $K^* = K(t \to \infty)$ are all negligibly small for very small values of $a$. With increasing $a$, $L^*$ and $P^*$ increase steadily and in a near-linear fashion, whereas $K^*$ suffers an abrupt high-hill jump at $a \sim 1$. Just beyond $a \sim 1$, the asymptotic keratinocyte density assumes the numerical value $K^* \approx 800$, which may be termed as the globally stable value of the keratinocyte density for the chosen set of model parameters. Results in the viewgraph indicates a threshold for leucocytes influx rate $a_{th} \sim 1.0$ beyond which disease pathogenesis in psoriasis sets in as manifested by the abrupt jump in $K^*$.

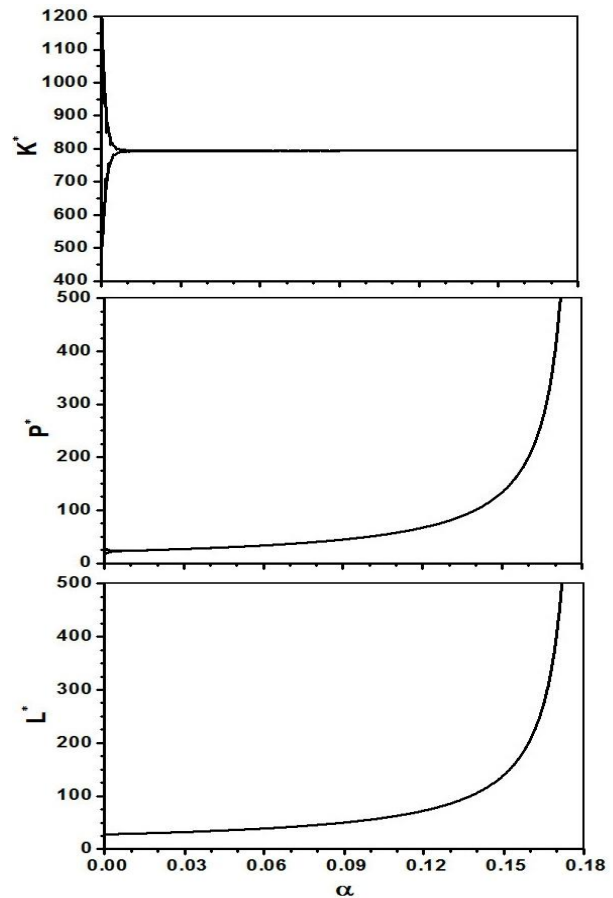

**Fig.6:** Asymptotic stable solutions of model variables $(L^*, P^*, K^*)$ with changing model parameter $\alpha$.

Phase diagram for varying model parameter $\alpha$, the growth-rate of leucocytes owing to proteins, has been included in Fig.6. Except varying the parameter $\alpha$, we keep all other parameters at their default values as in Table 1. We find that $L^*$ and $P^*$ increase slowly and monotonically at very small $\alpha$ and beyond some value of $\alpha \sim 0.12$, asymptotic stable solutions of leucocytes ($K^*$) and proteins ($P^*$) register almost exponential rise and diverges beyond $\alpha \sim 0.17$. Notice that for very small $\alpha$, $P^*$ solutions are multi-valued. Asymptotic solutions for keratinocytes' density $K^*$ is also multi valued at small $\alpha < 0.01$ bounded in a range 400-1200 (in mm$^{-3}$). However single valued saturation at $K^* \sim 800$ signifying a sharp disease pathogenesis beyond $\alpha_{th} \sim 0.01$.

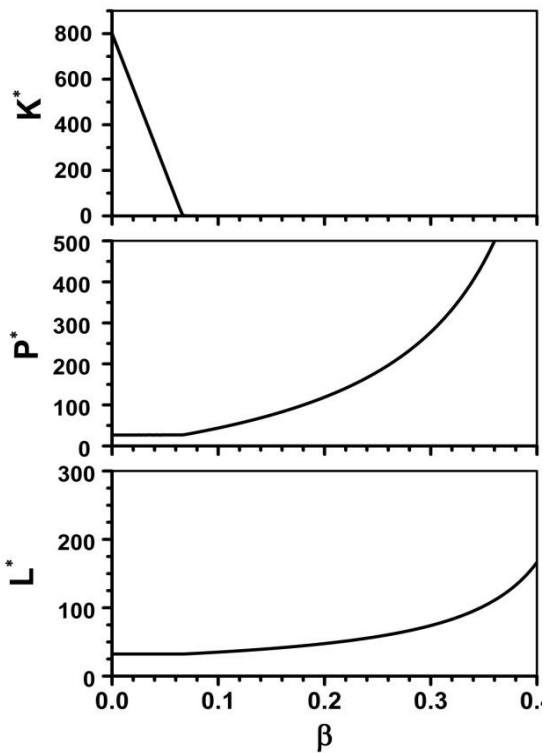

**Fig.7:** phase diagram involving asymptotic stable solutions of model variables $(L^*, P^*, K^*)$ with changing model parameter $\beta$.

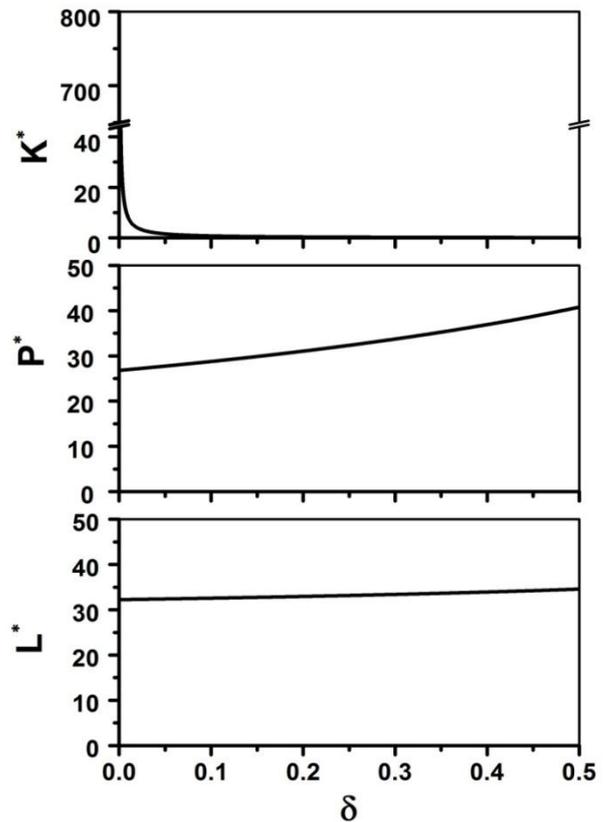

**Fig.8:** Asymptotic stable solutions of model variables $L^*, P^*$ and $K^*$ with changing model parameter $\delta$.

In Fig.7 we incorporate plots of $L^*$, $P^*$ and $K^*$ as functions of varying model parameter $\beta$, that gives the rate of growth of proteins due to leucocytes. At very small $\beta$, $L^*$ and $P^*$ are flat, but beyond $\beta \sim 0.08$, both leucocytes ($L^*$) and proteins ($P^*$) densities rise steeply. With increasing $\beta$, keratinocyte stable density $K^*$ drops linearly from its saturation value $\sim 800$ at a fast pace and becomes negligibly small beyond a threshold of $\beta_{th} \sim 0.065$. This viewgraph clearly indicates that at a moderate or large value of $\beta$ beyond the $\beta_{th}$, keratinocytes' density assumes very small numerical value with leucocytes ($L^*$) and proteins ($P^*$) densities being substantial. This Phenomenon points towards a sustainable roll-back from disease pathogenesis.

Fig.8 represents phase diagram of asymptotic model variable solutions with changing $\delta$ (protein growth rate owing to its mixing with keratinocytes). At very small $\delta$ (near zero), $L^*$ and $P^*$ are at their moderately small values whereas keratinocytes density is at its saturation $K^* \sim 800$. With increasing $\delta$, $L^*$ registers very little or almost no rise in its values while $P^*$ rises at low pace. However, increase in $\delta$ makes asymptotic keratinocytes' density $K^*$ to have a sharp fall. Thus we may assert that with increasing values of model parameter $\delta$ signature of pathogenesis in the disease psoriasis get smeared out.

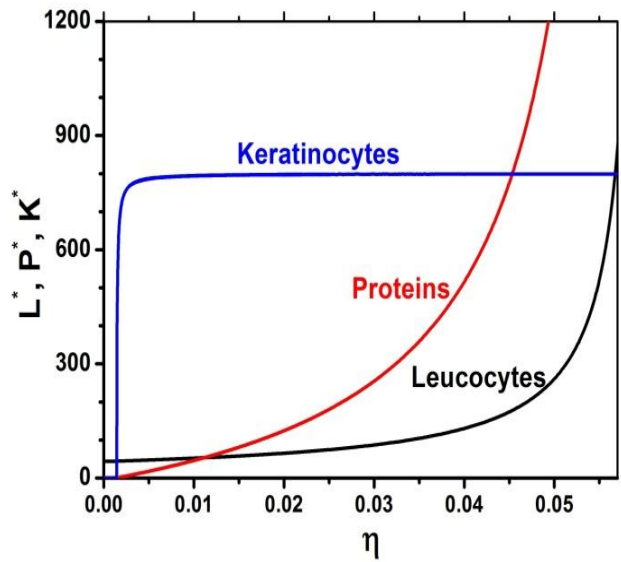

**Fig.9:** phase diagram involving asymptotic stable solutions of model variables $(L^*, P^*, K^*)$ with changing model parameter $\beta$.

Phase diagram of varying parameter $\eta$ is presented in Fig.9 where $\eta$ stands for growth rate of keratinocytes owing to its mixing with leucocytes. At very small $\eta$ ($< 0.015$), $L^*$ is small and $P^*$ & $K^*$ are negligibly small. With increasing $\eta$, asymptotic leucocyte density $L^*$ registers like-exponential rise and nearly diverges beyond $\eta \sim 0.055$ and asymptotic protein density $P^*$ rises faster than $L^*$ and also diverges beyond $\eta \sim 0.05$. Asymptotic keratinocytes' density $K^*$, from a very small value at small $\eta$, jumps to its globally stable saturation value $\sim 800$ at a threshold $\eta_{th} \sim 0.0015$. It can be asserted accordingly that disease pathogenesis occurs abruptly at a threshold $\eta_{th} \sim 0.0015$.

## 5 Conclusions and Further Work

The disease psoriasis, though manifested as chronic inflammation of skin, has its root to imbalances inflicted in the immune system of human (or any other mammalian) blood. As established by so far accumulated literature the disease is also termed as an autoimmune disease.

We, in this communication, developed a mathematical model for the autoimmune disease psoriasis based on the clinical and cell-biological evidences. The mathematical archetype has been framed on the buildup of a set of assumptions which are substantiated rightfully by clinical and cell-biological rulings gathered till date in the allied literature.

The mathematical model consists of three time differential equations involving pertinent cellular/biochemical masses that serve as model variables and these are connected to sequences of biological processes (or phenomena) being depicted in terms of a set of parameters symbolizing the strength of these biological processes or phenomena. The model, as a whole, actually provides population dynamic behaviors of variable masses of leucocytes, various proteins (cytokines, chemokines, growth factors) and the epidermal keratinocytes, where abnormally high proliferation of epidermal keratinocytes is implicative of disease pathogenesis. In our theoretical and numerical studies on the model, we have extensively explored effects of various model parameters on the dynamical progression of model variable masses and found that disease pathogenesis in psoriasis does exist within the model. Consequently, we also observe that interplay of some model parameters at specified strengths could signify sharp backtracking from disease pathogenesis.

With our analysis on the model we could establish that asymptotic solutions of the model are stable in the wide ranges of various model parameters. With the tool of numerical analysis being applied extensively in case of the present model, we could estimate (with considerable accuracy) numerical thresholds of some model parameters that inflict onset of or roll back from disease pathogenesis within the model. Note that, although we have embarked on the tempting practice of generating a set of default-valued model parameters. In actual practice, all the model parameters are varied in their respective permissible ranges capped by theoretical analysis.

We find that the model shows termination of disease pathogenesis when influx of leucocytes (signified by the strength of parameter $a$) is decreased below a threshold. Such roll-back from pathogenesis is analogous to T-cell (leucocytes) suppression by immuno-suppressive drugs improvised in clinical therapies. We further observe that an enhanced secretion of proteins by leucocytes, which is achieved within the model by increasing the parameter $\beta$, leads to a sustainable termination of disease pathogenesis and an increase in parameter $\delta$, representing the growth factor of proteins owing to its mixing with keratinocytes, also inflicts a smearing out of signatures of pathogenesis. These results are two important predictions from analysis of the present model which could be put to future clinical test.

Further work in the allied area would include a detailed justification of the default set of model parameters, exploration of added substantiating evidences (clinical and cell-biological) for the set of assumptions leading to the mathematical archetype and inclusion of realistic (parametric) functional dependence of biological processes involved within the model. We are also inclined to explore the possibility of applying our understanding of the present model in case of any other ailments pertaining to immune system.